\documentclass[a4paper]{article}

\usepackage{INTERSPEECH2020}
\usepackage{color}
\usepackage[table]{xcolor}
\usepackage{tipa}
\usepackage{booktabs}
\usepackage{multirow}
\usepackage{float}
\usepackage{flushend}
\usepackage{amsfonts,amsthm}

\PassOptionsToPackage{hyphens}{url}

\usepackage[bookmarks=false,hidelinks]{hyperref}

\hypersetup{
   unicode=false,          
   pdftoolbar=true,        
   pdfmenubar=true,        
   pdffitwindow=false,     
   pdfstartview={FitH},    
   pdftitle={My title},    
   pdfauthor={Author},     
   pdfsubject={Subject},   
   pdfcreator={Creator},   
   pdfproducer={Producer}, 
   pdfkeywords={keyword1} {key2} {key3}, 
   pdfnewwindow=true,      
   colorlinks=true,       
   linkcolor=blue,          
   citecolor=blue,        
   filecolor=magenta,      
   urlcolor=blue           
}

\usepackage{subfigure}
\usepackage{tikz}
\usepackage{pgfplots}
\pgfplotsset{compat=newest}
\usetikzlibrary{positioning,calc}
\usetikzlibrary{shapes,arrows}
\usepgfplotslibrary{groupplots}
\usepackage{amsmath,amssymb}
\usepackage{fdsymbol}
\usepackage{makecell}
\usepgfplotslibrary{statistics}
\renewcommand{\vec}[1]{\boldsymbol{\mathrm{#1}}}
\newcommand{\mtx}[1]{\boldsymbol{\mathrm{#1}}}
\newcommand{\transp}{\ensuremath{^\mathsf{T}}}

\newenvironment{customlegend}[1][]{%
	\begingroup
	\csname pgfplots@init@cleared@structures\endcsname
	\pgfplotsset{#1}%
}{%
\csname pgfplots@createlegend\endcsname
\endgroup
}%
\def\addlegendimage{\csname pgfplots@addlegendimage\endcsname}

\newcommand{\setvariable}[2]{
	\let#1\relax
	\newcommand{#1}{#2}
}

\usepackage{colorbrewer}

\usepackage{balance}

\colorlet{colorRaw}{BrBG-5-1}
\colorlet{colorDFTraw}{BrBG-5-2}
\colorlet{colorDFTsmoothed}{BrBG-5-2}
\colorlet{colorDFTcepstrum}{BrBG-5-2}
\colorlet{colorDFTsmoothed}{BrBG-5-2}
\colorlet{colorDFTcepstrum}{BrBG-5-2}
\colorlet{colorMixed}{BrBG-5-3}
\colorlet{colorCQTraw}{BrBG-5-4}
\colorlet{colorCQTcepstrum}{BrBG-5-4}
\colorlet{colorGroupDelay}{BrBG-5-5}
\colorlet{colorZTLCC}{BrBG-5-5}
\colorlet{colorOther}{BrBG-5-5}

\colorlet{colorDNN}{PRGn-5-1}
\colorlet{colorSEnet}{PRGn-5-1}
\colorlet{colorBayesianNN}{PRGn-5-1}
\colorlet{colorCNN}{PRGn-5-2}
\colorlet{colorLCNN}{PRGn-5-2}
\colorlet{colorResNet}{PRGn-5-2}
\colorlet{colorTDNN}{PRGn-5-3}
\colorlet{colorLSTM}{PRGn-5-4}
\colorlet{colorGMM}{PRGn-5-5}

\newcommand{\vspacePreCap}{\vspace{-0.5em}}
\newcommand{\vspacePostCap}{\vspace{-1em}}

\title{Visualizing Classifier Adjacency Relations:\\A Case Study in Speaker Verification and Voice Anti-Spoofing} 
\name{Tomi Kinnunen$^1$, Andreas Nautsch$^2$, Md Sahidullah$^3$, Nicholas Evans$^2$, Xin Wang$^4$,\\ Massimiliano Todisco$^2$, H\'ector Delgado$^5$, Junichi Yamagishi$^4$, Kong Aik Lee$^6$}

\address{
  $^1$University of Eastern Finland, Finland\\
  $^2$Eurecom, France\\
  $^3$Inria, France\\
  $^4$National Institute of Informatics, Japan\\
  $^5$Nuance Communications, Spain\\
  $^6$Institute for Infocomm Research, Singapore
}
\email{tomi.kinnunen@uef.fi,andreas.nautsch@eurecom.fr,md.sahidullah@inria.fr}

\begin{document}
\maketitle
\begin{abstract}
Whether it be for results summarization, or the analysis of classifier fusion, some means to compare different classifiers can often provide illuminating insight into their behaviour, (dis)similarity or complementarity. We propose a simple method to derive 2D representation from detection scores produced by an arbitrary set of binary classifiers in response to a common dataset. Based upon rank correlations, our method  facilitates a visual comparison of classifiers with arbitrary scores and with close relation to receiver operating characteristic (ROC) and detection error trade-off (DET) analyses. While the approach is fully versatile and can be applied to any detection task, we demonstrate the method using scores produced by automatic speaker verification and voice anti-spoofing systems. The former are produced by a Gaussian mixture model system trained with VoxCeleb data whereas the latter stem from submissions to the ASVspoof 2019 challenge.  
\end{abstract}
\noindent\textbf{Index Terms}: classifier, multi-dimensional scaling

\section{Introduction} 

Whether it be for challenge results summarization~\cite{Nautsch2021-ASVspoof19-countermeasures,lee2020-two-decades,Greenberg2020-two-decades,Nagrani2020-voxceleb} or the analysis of classifier fusion, some means to compare alternative classifiers can often provide illuminating insight into robustness, coherence, and generalisation \cite{Ramos-LRvalidation-2017}; (dis)similarity, and complementarity. Depending on prior knowledge (and data) available, one could compare
    \begin{itemize}
        \item the underlying working principles, 
        features, architectures, training data properties --- \textbf{classifier metadata},
        \item \textbf{empirical performance} (e.g.\ accuracy) on a common set of evaluation data,
    \end{itemize}
to name two possibilities. Often we want to understand how the two are related --- what is the impact of specific modeling choices upon performance. A common approach is to fix some parameters while varying others to find out their impact on performance. Whenever one implements the classifier and has full control over the experiment, this \emph{white-box} approach is relatively straightforward. There are, however, situations where one may not have access to exhaustive classifier details and configuration settings but would still like to learn about similarities between alternative solutions. One such \emph{black-box} scenario are public machine learning challenges and technology benchmarks. Whether it be NIST speaker recognition evaluations~\cite{lee2020-two-decades,Greenberg2020-two-decades}, VoxCeleb \cite{Nagrani2020-voxceleb}, ASVspoof~\cite{Nautsch2021-ASVspoof19-countermeasures}, or indeed any other competitive campaigns, participants typically run their in-house systems on a common evaluation set and submit predictions (e.g.\ scores) for unlabeled data along with a system description. Since source codes or models are often not required, there can be uncertainty as to why specific challenge entries outperform others. The authors' personal motivation for the presented work stems partially from difficulties encountered in our efforts to link classifier properties to their performance in 
a recent challenge~\cite{Nautsch2021-ASVspoof19-countermeasures}. We wanted to find out what can be learnt about classifier differences based on detection scores.

\begin{figure*}[!t]
    \centering
    \includegraphics[width=12.8cm]{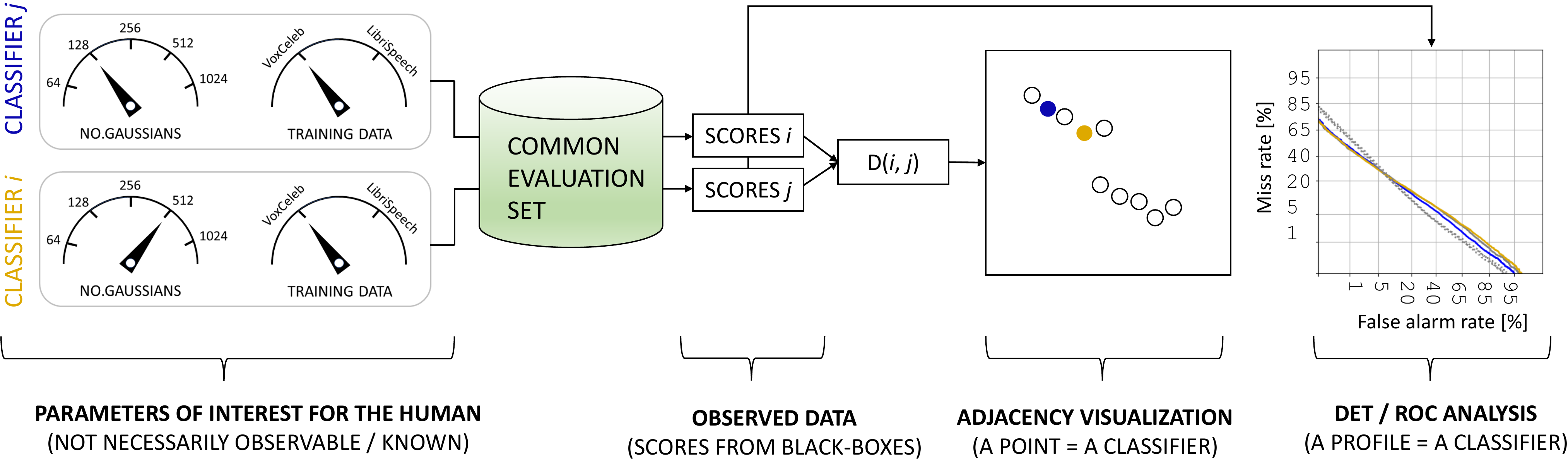}\vspacePreCap
    \caption{The behavior of classifiers can be controlled using a limited number of parameters meaningful to the experimenter. Not all the details, however, may be available to an evaluator. We propose to visualize classifier adjacencies based on their scores to complement other traditional analyses. The adjacencies can be informative of the adjacency of the underlying control parameters.\vspacePostCap}
    \label{fig:method-idea}
\end{figure*}

To this end, we propose a method for computing the distance between binary classifiers based on the scores produced for common evaluation data (Fig. \ref{fig:method-idea}). We impose a stronger notion of classifier (dis)similarity than the established methods of \emph{detection error trade-off} (DET) \cite{Martin97-DET-curve} and \emph{receiver operating characteristics} (ROC) analyses: to be considered identical, a pair of classifiers must agree not only in their DET profiles but relative ordering of individual trials. Similar principles are the basis for a number of statistical 
significance tests but our perspective is in visualizing classifiers beyond 
DET/ROC plots.

As Fig.~\ref{fig:method-idea} suggests, we conjecture that distances derived from scores may provide information on (possibly unknown) classifier metadata differences. After describing the background and methodology, we report an example application of our methodology to automatic speaker verification (ASV) and voice anti-spoofing. The former includes classifiers constructed by the authors (known classifier metadata) while the latter represents the less controlled case of submissions to the ASVspoof 2019 challenge. Note, however, that the proposed methods are not in any way specific or limited to the speech domain. To this end, we provide an open-source reference implementation\footnote{\url{https://github.com/asvspoof-challenge/classifier-adjacency} (referenced June 10, 2021).}.

\section{Trade-offs in Characterizing Classifiers}

There are various ways to describe a classifier. The coarsest description might include the general model class --- such as  `a linear classifier' or `a deep neural network.' A more refined description might detail the features and their dimensionality, the type and number of layers, the software toolkit, loss function, or training corpus to name a few possibilities. 

Depending on the target group (e.g.\ general public, scientists, students, software engineers) one may prefer different levels of detail. What is in common is that the descriptions are designed for \emph{communicating} key principles rather than enabling perfect reproducibility. The description is a \underline{h}uman-readable tuple, $\vec{h}$. For instance, our automatic speaker verification (ASV) systems described below are characterized by triplets $\vec{h}=(\texttt{TrData},F,G)$ where $\texttt{TrData} \in \{\texttt{VoxCeleb}, \texttt{LibriSpeech}\}$ indicates the training corpus while $F$ and $G$ are integers that indicate the number of mel filters and Gaussians, respectively. Each choice (attribute value in $\vec{h}$) impacts results but there is no expectation for such triplets to be a complete description; it is a description intended for a particular audience (here, Interspeech attendee).

At the other extreme, the most precise description is a packaged/trained model --- a list of all  parameters, $\vec{\theta} \in\mathbb{R}^D$. For instance, $\vec{\theta}$ could represent all the weights of a deep neural network (DNN), $D$ being potentially in the order of millions. If scoring is deterministic (as is usually the case), $\vec{\theta}$ and the functional form $g(X;\vec{\theta})$ completely specify how an arbitrary input $X$ maps to an output (predicted class label or score), $X \mapsto \hat{Y}=g(X;\vec{\theta})$. This leads to perfect reproducibility --- one obtains the same scores, and hence also results, every time. Unlike $\vec{h}$, however, $\vec{\theta}$ is not intended to be comprehensible to the human: eyeballing the numerical values of DNN weights (an exercise which could take a while) is not particularly helpful in understanding \emph{how} the system works or \emph{why} it may work/fail in specific cases. Another problem is incomparability of the parameter vectors of different models. They could have different dimensionalities and there may be no meaningful correspondence between dimensions. 

One more level of description are the scores that can be used to derive detection error trade-off (DET) profiles~\cite{Martin97-DET-curve} and metrics such as the equal error rate (EER). Unfortunately, being based on score \emph{distributions}, DETs or EERs are not informative of classifier responses to individual trials. A common compromise is to define a limited number of \emph{evaluation conditions} --- subsets of trials with characteristics useful for diagnostics (e.g.\ `telephony', `short-duration', `attack S10', `females'). Performance breakdown by condition is useful in analysing trends.

In summary, there are different abstraction levels to characterize classifiers. We aim at learning about their differences in terms of human-targeted descriptions \emph{without the precise knowledge of the attribute values themselves}. We conjecture that classifiers that are similar in their $\vec{h}$ descriptions may produce similar outputs when executed on common data. In this case, one can learn about classifier design differences from their scores. To this end, we propose a method to compute pairwise classifier distances 
for common evaluation data. The distances can then be used for different purposes, such as agglomerative clustering or visualisation, to help in demonstrating trends or diagnosing classifier complementarities.

\section{Proposed Classifier Adjacency Visualizer}\label{sec:methodology}

\subsection{Assumed input data}

Let $X$ and $Y$ denote some input space and (binary) labels, respectively. We use $\mathcal{D}=\{(X_i,Y_i)\}_{i=1}^N \sim_\text{i.i.d.} p(X,Y)$ to denote a labeled (supervised) evaluation set of $N$ \emph{trials}. In an ASV set-up, each $X_i$ corresponds to an enrollment-test pair and $Y_i \in \{0,1\}$ indicates whether the speaker identities in the two utterances are the same (target trial) or different (nontarget trial). In voice anti-spoofing, in turn, each $X_i$ corresponds to a test file and $Y_i$ indicates whether $X_i$ represents bonafide audio or a spoofing attack (such as computer-generated speech).

We assume a set  $\mathcal{G}=\{g_1,\dots,g_M\}$ of $M$ \emph{binary classifiers} (detectors), $g_j: X \rightarrow \mathbb{R}$, each of which assigns a numeric membership value (\emph{detection score}) to each trial, with the convention that high scores indicate support towards the positive class (e.g.\ target speaker). Unless we construct the classifiers ourselves, we do not have access to the scoring functions $\{g_j\}$ but only to their \emph{outputs} for the shared evaluation data $\mathcal{D}$ (in turn, in a challenge setting, only the organizer knows the ground truth values $\{Y_i\}$ while the participants produce predictions on the test trials $\{X_i\}$ blindly). We use $s_{i,j}=g_j(X_i)$ to denote the response of the $j$th classifier to the $i$th trial. The data used for our modeling consists of the responses of all classifiers to all the test trials --- an $N \times M$ matrix $\mtx{S}=[\vec{s}_1,\dots,\vec{s}_M]$ where $\vec{s}_j=[s_{1,j},\dots,s_{N,j}]\transp$ is a column vector of size $N \times 1$. 

Optionally, one may have additional trial metadata. In this case, we assume that each trial can be uniquely assigned to one member of a  mutually exclusive set of \emph{groups} (conditions), $c \in \{1,2,\dots,C\}$. In our experiments, for instance, groups in the ASV application correspond to the specific pair of speaker identities in the trial. We denote the number of trials in group $c$ by $N_c$. As the groups are mutually exclusive, $N = \sum_{c=1}^{C} N_c$. We use either the original $N \times M$ score matrix or the smaller $C \times M$ matrix $\mtx{S'}=[\vec{\mu}_1,\dots,\vec{\mu}_M]$ where each group is represented using its average score. Here, $\vec{\mu}_j = [\mu_{1,j},\dots,\mu_{C,j}]\transp$ is a column vector of size $C \times 1$ for the $j$th classifier.

\subsection{Classifier Adjacency}

To identify similar (and dissimilar) systems, we define a distance function between classifiers, indicated by $D(i,j)$, where $i, j \in \{1,\dots,M\}$. It is computed either from $\vec{S}$ or $\vec{S}'$. We first note that, since the numerical range of detection scores can be arbitrary, a direct comparison of scores (e.g.\ using Euclidean distance) is usually not meaningful. In the ASV field, `correction' for the range of scores is typically addressed through \emph{calibration} (e.g.~\cite{Brummer-deVilliers-BOSARIS-Binary-Scores-AGNITIO-Research-2011}) that converts arbitrary scores into calibrated log-likelihood ratios (LLRs). Here, however, since score calibration is not our main concern, we opt for more straightforward scale- and translation-invariant distance computation. To this end, we define classifier distance through \emph{order statistics} as opposed to raw scores. This choice stems from the knowledge that any two classifiers which yield the same order of trial scores on common data will share the same detection error trade-off~(DET) curve and equal error rate~(EER)~\cite[p.81]{brummer2010-phd}\footnote{The converse does not necessarily hold: having the same DET curve or EER does not imply that the trial rankings must be the same.}.

We use \emph{Kendall's} $\tau$~\cite{kendall1938measure,Kendall1945-tie-treatment,knight66computer} as the rank correlation measure between classifiers $i$ and $j$: 
\renewcommand{\theequation}{2}
    \begin{equation}\label{eq:kendall-tau}
        \tau(i,j) = \frac{N_\text{con}-N_\text{dis}}{\sqrt{(N_\text{con} + N_\text{dis} + T_i)\times (N_\text{con} + N_\text{dis} + T_j)}},
    \end{equation}
where $N_\text{con}$ and $N_\text{dis}$ are, respectively, the number of concordant and discordant pairs between $i$ and $j$. Further, $T_i$ is the number of ties only in $i$, and $T_j$ is the number of ties only in $j$. \emph{Concordance} of a pair of trials means that the sort order agrees across the classifiers. For instance, if the $a^\text{th}$ and $b^\text{th}$ trial scores are ordered $s_{a,i} < s_{b,i}$ in classifier $i$, the trial pair $(a,b)$ is cordordant with classifier $j$, if similarly $s_{a,j} < s_{b,j}$. Likewise, if $(s_{a,i} > s_{b,i})$ \texttt{AND} $(s_{a,j} > s_{b,j})$, pair $(a,b)$ is again corcondant. If concordance is not satisfied, the pair $(a,b)$ is \emph{discordant}. A \emph{tie} within either classifier means that there identical scores produced for different trials. With real-valued scores stored in float or double precision this is generally a rare case.

An intuitive feeling of \eqref{eq:kendall-tau} can be developed by first assuming there are no ties ($T_i=T_j=0$). Now, if the two classifiers place all trials into the same sort order, $N_\text{dis}=0$ and $\tau(i,j)$ reaches the maximum value of $1$. The other extreme is obtained when the sort order of trials of one classifier is the reverse of the other. This corresponds to  $N_\text{con}=0$ and yields  the minimum value of $-1$. 
Finally, $\tau=0$ indicates lack of statistical association between the two. To sum up, $\tau(i,j)$ takes values in $[-1,1]$ (similar to Pearson correlation) and can be considered as a degree of agreement in the sort order of trials. Whenever $\tau(i,j)=1$, the classifiers have identical DET or ROC profiles. 

For the purposes of visualization, we map $\tau$ into distances as $D(i,j)=\frac{1}{2}(1-\tau(i,j))$. Hence, identical systems (in the sense of their Kendall $\tau$) map to a distance of 0 while reverse ordering yields a maximum distance of~1. Finally, visualizations of classifier adjacency relations are obtained using classical (non-metric) \emph{multidimensional scaling}~(MDS). Each classifier is represented by a point in 2D space so that between-classifier distances approximate those in the given distance matrix (here, the $M \times M$ matrix containing all pairwise distances).

\section{Experimental set-up}

We demonstrate the proposed methods with two different tasks: automatic speaker verification (ASV) and voice anti-spoofing. The former consists of classifiers constructed by the authors with controlled parameters. The latter consists of countermeasures submitted to the ASVspoof 2019 challenge. Even if the system descriptions are known, we cannot interact with these systems nor do we know all their implementation nuances. 

\subsection{Automatic speaker verification}
The analysis of ASV methods uses the publicly available VoxCeleb dataset. We use the standard trial list of 40 test speakers in VoxCeleb1~\cite{nagrani17voxceleb}. It consists of 18860 target trials and the same number of nontarget trials. We form the groups (see Section~\ref{sec:methodology}) from the cross-product of speaker identities in the trials. With 40 speakers, this yields $(40 \cdot 41)/2=820$ groups. 

We use Gaussian mixture model-universal background model (GMM-UBM) based ASV system with an MFCCs front-end. While there are  better ASV methods, we opt for the computationally light GMM-UBM as we want to generate a large number of classifiers. The focus of this study is in classifier adjacency rather than performance. 

Following standard practice, extracted MFCCs are augmented with delta features and processed with RASTA and utterance-level \emph{cepstral mean and variance normalization} (CMVN). For MFCCs, we include filterbanks comprising 12, 16, 20 and 24 filters. For UBMs, we vary the number of components in the powers of 2 between 2 and 1024. For UBM training, we select two different datasets: VoxCeleb2~\cite{Nagrani2020-voxceleb} and LibriSpeech~\cite{panayotov2015librispeech}. In total, we have 80 ASV systems.

\setvariable{\mdsheight}{4.0cm}
\setvariable{\mdswidth}{5.4cm}
\begin{figure}[t]
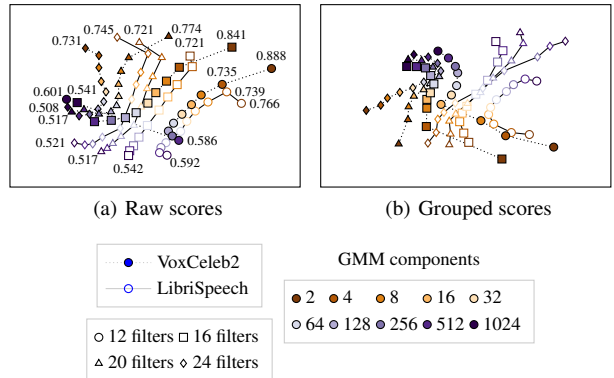

    \centering
    \subfigure[Raw scores]{\input{figs/asv.tikz}}
    \hfill
    \subfigure[Grouped scores]{\input{figs/asv_mean.tikz}}\\
    \begin{tikzpicture}[font=\scriptsize]
    \node (db) at (0,0) {
    \begin{tikzpicture}[font=\scriptsize]
        \begin{customlegend}[
            legend cell align=left,
            legend columns=1,
            legend entries={
                VoxCeleb2, LibriSpeech
            },
            legend style={draw=white!80!black,anchor=north}]
                \addlegendimage{mark=*, densely dotted,mark options={scale=0.75,solid,draw=black,thin,fill=blue}}
                \addlegendimage{mark=o, solid, mark options={scale=0.75,draw=blue,thin,fill=white}}
        \end{customlegend}
    \end{tikzpicture}};
    \node[below=-.25em of db.south] (filter) {
    \begin{tikzpicture}[font=\scriptsize]
        \begin{customlegend}[
            legend cell align=left,
            legend columns=2,
            legend entries={
                12 filters, 16 filters, 20 filters, 24 filters
            },
            legend style={draw=white!80!black,anchor=north}]
                \addlegendimage{only marks ,mark=o, mark options={scale=0.75,draw=black,thin,fill=white}}
                \addlegendimage{only marks ,mark=square, mark options={scale=0.75,draw=black,thin,fill=white}}
                \addlegendimage{only marks ,mark=triangle, mark options={scale=0.75,draw=black,thin,fill=white}}
                \addlegendimage{only marks ,mark=diamond, mark options={scale=0.75,draw=black,thin,fill=white}}
        \end{customlegend}
    \end{tikzpicture}};
    \node[right=-.25em of filter.east,yshift=1.625em,align=center] (mix) {
    \begin{tikzpicture}[font=\scriptsize]
        \begin{customlegend}[
            legend cell align=left,
            legend columns=5,
            legend entries={
                2, 4, 8, 16, 32, 64, 128, 256, 512, 1024 
            },
            legend style={draw=white!80!black,anchor=north}]
                \addlegendimage{only marks ,mark=*, mark options={scale=0.75,draw=black,thin,fill=PuOr-10-1}}
                \addlegendimage{only marks ,mark=*, mark options={scale=0.75,draw=black,thin,fill=PuOr-10-2}}
                \addlegendimage{only marks ,mark=*, mark options={scale=0.75,draw=black,thin,fill=PuOr-10-3}}
                \addlegendimage{only marks ,mark=*, mark options={scale=0.75,draw=black,thin,fill=PuOr-10-4}}
                \addlegendimage{only marks ,mark=*, mark options={scale=0.75,draw=black,thin,fill=PuOr-10-5}}
                \addlegendimage{only marks ,mark=*, mark options={scale=0.75,draw=black,thin,fill=PuOr-10-6}}
                \addlegendimage{only marks ,mark=*, mark options={scale=0.75,draw=black,thin,fill=PuOr-10-7}}
                \addlegendimage{only marks ,mark=*, mark options={scale=0.75,draw=black,thin,fill=PuOr-10-8}}
                \addlegendimage{only marks ,mark=*, mark options={scale=0.75,draw=black,thin,fill=PuOr-10-9}}
                \addlegendimage{only marks ,mark=*, mark options={scale=0.75,draw=black,thin,fill=PuOr-10-10}}
        \end{customlegend}
    \end{tikzpicture}};
    \node[above=-0.25em of mix] {GMM components};
\end{tikzpicture}\vspacePreCap
    \caption{Visualisation of ASV similarity on VoxCeleb1 test set. The ASV systems are toy classifiers to show case how the proposed visualisation method behaves in a controlled setting. Shown $C_\textit{llr}^\textit{min}$ values serve nominal orientation.\vspacePostCap}
    \label{fig:asv-mds}
\end{figure}

\subsection{Anti-spoofing}

The analysis of voice anti-spoofing methods, or \emph{spoofing countermeasures} (CMs), uses ASVspoof 2019 challenge submission entries. Details of the dataset \cite{Wang2020-ASVspoof-database} and challenge results \cite{Nautsch2021-ASVspoof19-countermeasures} are provided elsewhere. Here we focus on aspects relevant for the novel classifier adjacency analysis.

The challenge data consists of two different scenarios. The evaluation set of the \emph{logical attack} (LA) scenario contains 13 different text-to-speech or voice conversion attacks (labeled A07$\dots$A19). The physical attack (PA) scenario, in turn, consists of simulated replay attacks from 27 different environments and 9 replay configurations. Both sets also contain additional \emph{bona fide} (human speech) utterances. The participants processed the corresponding audio files blindly to obtain detection scores. In the LA scenario, the 3-class evaluation set consists of $5\,370$ bonafide/target, $1\,985$ bonafide/non-target, and $63\,882$ spoofed trials. For the PA scenario, the corresponding numbers are $12\,960$ bonafide/target, $5\,130$ bonafide/non-target, and $116\,640$ spoofed trials.  For the LA scenario, the groups (see Section \ref{sec:methodology}) are formed from the 13 attacks plus the bonafide class (a total of 14 groups). For the PA scenario, we used all cross-combinations of 27 environments $\times$ (9 replay configurations + 1 bona fide) to define 270 groups.

\section{Results}

The proposed method is demonstrated in a controlled-classifier setting (ASV) and in a setting of unconstrained classifiers (CMs).

\subsection{Speaker Verification}

Fig.~\ref{fig:asv-mds} illustrates ASV results with varying number of filters, mixtures, and training data. The indicated $C_\text{llr}^\text{min}$ performance~\cite{Brummer-deVilliers-BOSARIS-Binary-Scores-AGNITIO-Research-2011,brummer2006application} is demonstrated for the lowest and highest number of mixture components to provide nominal guidance only. For the top-5 systems, $C_\text{llr}^\text{min}$ varies between 0.466 and 0.481 (for visual clarity, we do not show all the $C_\text{llr}^\text{min}$ values in Fig. ~\ref{fig:asv-mds}).

Differences in classifier parameters expose observable trends. For 12 filters, decision boundaries resulting from different training data become closer for as few as $4$ components. Similar trajectories are followed as the number of components increases (as does the discrimination performance). Then, while the LibriSpeech trajectory continues, the VoxCeleb trajectory makes a sudden jump into the area of the top classifiers (at 512/1024 components and 20/24 filters). Adjacent classifiers (to the top ones) have less filters but more components; the top classifiers use more filters but less components. 
Our visuals supplement conventional DET plots: they provide insights into how decision boundaries are affected by data and parameters.

\setvariable{\mdswidth}{3.4cm}
\setvariable{\mdsheight}{3.4cm}
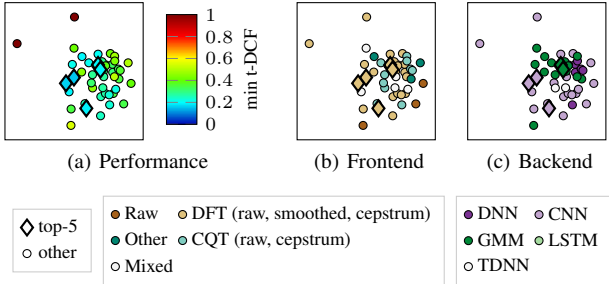
\begin{figure}[t]
    \centering
    \subfigure[Performance]{        \begin{tikzpicture}[font=\scriptsize]
            \begin{axis}[
            width=\mdswidth,
            height=\mdsheight,
            ticks=none,
            xmin=-0.763361, xmax=0.378251,
            ymin=-0.438021, ymax=0.51552,
            colormap/bluered,
            colorbar,
            colorbar style={scale=0.8,yshift=-0.5em,ylabel={min t-DCF},
                ylabel style={yshift=0.5ex}
                },
            scatter/use mapped color={draw=black,fill=mapped color},
            point meta=explicit,
            point meta min=0.0,
            point meta max=1.0
            ]
                
\addplot[scatter, only marks, mark=*, mark options={scale=0.75}]
coordinates{
(0.005607, 0.157255) [0.2077]
(0.028692000000000002, -0.138207) [0.2586]
(-0.070867, 0.097349) [0.2106]
(0.254181, 0.097204) [0.4971]
(0.10694200000000001, 0.161856) [0.413]
(-0.130219, 0.032997000000000005) [0.4451]
(0.070123, 0.082679) [0.3819]
(-0.039484, -0.07774199999999999) [0.4508]
(-0.049904000000000004, -0.042359) [0.3143]
(-0.663361, 0.23063499999999998) [1.0]
(-0.218268, -0.338021) [0.549]
(0.101528, -0.19510999999999998) [0.2305]
(0.12636, 0.030799) [0.3057]
(0.13167, -0.116147) [0.3456]
(0.217245, -0.188436) [0.3415]
(-0.013819, 0.080457) [0.2124]
(0.039159, 0.10505) [0.2865]
(-0.18913, 0.41552) [1.0]
(0.124395, 0.08667000000000001) [0.3237]
(-0.127095, 0.140438) [0.1949]
(0.026378, -0.271111) [0.246]
(0.178557, -0.015436000000000002) [0.3527]
(0.174191, 0.053774) [0.3294]
(0.14064100000000002, 0.0005009999999999999) [0.5131]
(0.278251, -0.045966) [0.3964]
(-0.193048, 0.198871) [0.4079]
(0.066771, -0.116004) [0.2653]
(0.188661, 0.044331999999999996) [0.3632]
(-0.216969, -0.22390100000000002) [0.3482]
(0.045072, -0.076382) [0.273]
(-0.233724, -0.10441099999999999) [0.1946]
(-0.040445999999999996, 0.014316999999999998) [0.284]
(0.151666, -0.08657100000000001) [0.2202]
(0.049983, 0.038271) [0.3135]
(0.142549, 0.143077) [0.474]
(-0.103102, -0.1866) [0.2555]
(-0.20681799999999997, 0.08608099999999999) [0.2128]
(0.052689, 0.000367) [0.2774]
(0.097419, -0.12548900000000002) [0.23]
(0.06797, 0.076593) [0.3845]
(0.143572, 0.10993299999999999) [0.4346]
};

\addplot[scatter,only marks,mark=diamond*, mark options={scale=1.5, thick}]
coordinates{
(0.005611, 0.08254600000000001) [0.1937]
(0.024919999999999998, 0.050095999999999995) [0.1939]
(-0.259402, -0.045491000000000004) [0.1655]
(-0.091473, -0.21937199999999998) [0.1894]
(-0.193675, -0.004911) [0.1562]
};

        \end{axis}
        \end{tikzpicture}}
    \hfill
    \subfigure[Frontend]{\begin{tikzpicture}[font=\scriptsize]
    \begin{axis}[
    width=\mdswidth,
    height=\mdsheight,
    ticks=none,
    xmin=-0.763361, xmax=0.378251,
    ymin=-0.438021, ymax=0.51552
    ]
        
\addplot[only marks, mark=*, mark options={scale=0.75,solid, fill=BrBG-5-1}]
table{%
x                      y
-0.218268 -0.338021
0.217245 -0.188436
0.278251 -0.045966
};

\addplot[only marks, mark=*, mark options={scale=0.75,solid, fill=BrBG-5-2}]
table{%
x                      y
0.005607 0.157255
0.028692000000000002 -0.138207
0.10694200000000001 0.161856
-0.130219 0.032997000000000005
0.070123 0.082679
-0.663361 0.23063499999999998
0.12636 0.030799
0.13167 -0.116147
-0.013819 0.080457
0.039159 0.10505
-0.18913 0.41552
0.124395 0.08667000000000001
0.026378 -0.271111
0.14064100000000002 0.0005009999999999999
0.066771 -0.116004
-0.216969 -0.22390100000000002
0.049983 0.038271
-0.20681799999999997 0.08608099999999999
0.097419 -0.12548900000000002
0.06797 0.076593
};

\addplot[only marks, mark=*, mark options={scale=0.75,solid, fill=BrBG-5-3}]
table{%
x                      y
-0.193048 0.198871
0.045072 -0.076382
-0.233724 -0.10441099999999999
0.151666 -0.08657100000000001
0.052689 0.000367
};

\addplot[only marks, mark=*, mark options={scale=0.75,solid, fill=BrBG-5-4}]
table{%
x                      y
-0.070867 0.097349
-0.039484 -0.07774199999999999
-0.049904000000000004 -0.042359
0.101528 -0.19510999999999998
0.174191 0.053774
0.188661 0.044331999999999996
-0.040445999999999996 0.014316999999999998
0.142549 0.143077
-0.103102 -0.1866
0.143572 0.10993299999999999
};

\addplot[only marks, mark=*, mark options={scale=0.75,solid, fill=BrBG-5-5}]
table{%
x                      y
0.254181 0.097204
-0.127095 0.140438
0.178557 -0.015436000000000002
};

\addplot[only marks, mark=diamond*, mark options={scale=1.5,solid, thick,fill=BrBG-5-2}]
table{%
x                      y
0.005611 0.08254600000000001
0.024919999999999998 0.050095999999999995
-0.259402 -0.045491000000000004
-0.091473 -0.21937199999999998
-0.193675 -0.004911
};

\end{axis}
\end{tikzpicture}}
    \hfill
    \subfigure[Backend]{        \begin{tikzpicture}[font=\scriptsize]
            \begin{axis}[
            width=\mdswidth,
            height=\mdsheight,
            ticks=none,
            xmin=-0.763361, xmax=0.378251,
            ymin=-0.438021, ymax=0.51552
            ]
                
\addplot[only marks, mark=*, mark options={scale=0.75,solid, fill=PRGn-5-1}]
table{%
x                      y
0.101528 -0.19510999999999998
0.12636 0.030799
0.124395 0.08667000000000001
0.174191 0.053774
0.188661 0.044331999999999996
0.143572 0.10993299999999999
};

\addplot[only marks, mark=*, mark options={scale=0.75,solid, fill=PRGn-5-2}]
table{%
x                      y
0.005607 0.157255
0.028692000000000002 -0.138207
0.254181 0.097204
0.10694200000000001 0.161856
-0.049904000000000004 -0.042359
-0.663361 0.23063499999999998
0.13167 -0.116147
0.217245 -0.188436
-0.18913 0.41552
0.026378 -0.271111
0.278251 -0.045966
0.066771 -0.116004
-0.216969 -0.22390100000000002
-0.233724 -0.10441099999999999
0.151666 -0.08657100000000001
-0.103102 -0.1866
0.052689 0.000367
0.097419 -0.12548900000000002
};

\addplot[only marks, mark=*, mark options={scale=0.75,solid, fill=PRGn-5-3}]
table{%
x                      y
-0.039484 -0.07774199999999999
0.045072 -0.076382
};

\addplot[only marks, mark=*, mark options={scale=0.75,solid, fill=PRGn-5-5}]
table{%
x                      y
-0.070867 0.097349
-0.130219 0.032997000000000005
0.070123 0.082679
-0.218268 -0.338021
-0.013819 0.080457
0.039159 0.10505
-0.127095 0.140438
0.178557 -0.015436000000000002
0.14064100000000002 0.0005009999999999999
-0.193048 0.198871
-0.040445999999999996 0.014316999999999998
0.049983 0.038271
0.142549 0.143077
-0.20681799999999997 0.08608099999999999
0.06797 0.076593
};

\addplot[only marks, mark=diamond*, mark options={scale=1.5,solid, thick,fill=PRGn-5-1}]
table{%
x                      y
};

\addplot[only marks, mark=diamond*, mark options={scale=1.5,solid, thick,fill=PRGn-5-2}]
table{%
x                      y
-0.259402 -0.045491000000000004
-0.091473 -0.21937199999999998
-0.193675 -0.004911
};

\addplot[only marks, mark=diamond*, mark options={scale=1.5,solid, thick,fill=PRGn-5-3}]
table{%
x                      y
};

\addplot[only marks, mark=diamond*, mark options={scale=1.5,solid, thick,fill=PRGn-5-5}]
table{%
x                      y
0.005611 0.08254600000000001
0.024919999999999998 0.050095999999999995
};

        \end{axis}
        \end{tikzpicture}}\\
    \begin{tikzpicture}[font=\scriptsize]
    \node (topother) at (0,0) {
    \begin{tikzpicture}[font=\scriptsize]
    \begin{customlegend}[
        legend cell align=left,
        legend columns=1,
        legend entries={
            top-5,
            other
        },
        legend style={draw=white!80!black,anchor=north}]
        \addlegendimage{only marks,mark=diamond*, mark options={scale=1.5,fill=white,solid, thick,draw=black}}
        \addlegendimage{only marks,mark=*, mark options={scale=0.75,fill=white,draw=black}}
    \end{customlegend}\end{tikzpicture}};
    \node[right=-0.25em of topother.east] (frontend) {
    \begin{tikzpicture}[font=\scriptsize]
    \begin{customlegend}[
        legend cell align=left,
        legend columns=2,
        legend entries={
            Raw,
            {DFT (raw, smoothed, cepstrum)},
            Other,
            {CQT (raw, cepstrum)},
            Mixed
        },
        legend style={draw=white!80!black,anchor=north}]
        \addlegendimage{only marks ,mark=*, mark options={scale=0.75,draw=black,fill=colorRaw}}
        \addlegendimage{only marks ,mark=*, mark options={scale=0.75,draw=black,fill=colorDFTraw}}
        \addlegendimage{only marks ,mark=*, mark options={scale=0.75,draw=black,fill=colorOther}}
        \addlegendimage{only marks ,mark=*, mark options={scale=0.75,draw=black,fill=colorCQTraw}}
        \addlegendimage{only marks ,mark=*, mark options={scale=0.75,draw=black,fill=colorMixed}}
    \end{customlegend}\end{tikzpicture}};
    \node[right=-0.25em of frontend.east] (backend) {
    \begin{tikzpicture}[font=\scriptsize]
    \begin{customlegend}[
        legend cell align=left,
        legend columns=2,
        legend entries={
            DNN,
            CNN,
            GMM,
            LSTM,
            TDNN
        },
        legend style={draw=white!80!black,anchor=north}]
        \addlegendimage{only marks ,mark=*, mark options={scale=0.75,draw=black,fill=colorDNN}}
        \addlegendimage{only marks ,mark=*, mark options={scale=0.75,draw=black,fill=colorCNN}}
        \addlegendimage{only marks ,mark=*, mark options={scale=0.75,draw=black,fill=colorGMM}}
        \addlegendimage{only marks ,mark=*, mark options={scale=0.75,draw=black,fill=colorLSTM}}
        \addlegendimage{only marks ,mark=*, mark options={scale=0.75,draw=black,fill=colorTDNN}}
    \end{customlegend}\end{tikzpicture}};
    \end{tikzpicture}\vspacePreCap
    \caption{CM similarity on ASVspoof 2019 LA test set. The CM systems are single systems (no fusions) submitted by participants; an uncontrolled setting is showcased (raw scores). The ASVspoof 2019 primary metric is shown, the so-called \emph{min t-DCF} \cite{Kinnunen2020_tDCF}. Colours show different frontend transforms \& feature types and backend classifiers.\vspacePostCap}
    \label{fig:mds-cm-la}
\end{figure}

\subsection{Anti-spoofing}

\begin{figure}[t]
    \centering
    \subfigure[Raw/FE]{\begin{tikzpicture}[font=\scriptsize]
    \begin{axis}[
    width=\mdswidth,
    height=\mdsheight,
    ticks=none,
    xmin=-0.368131, xmax=0.543509,
    ymin=-0.383389, ymax=0.65787
    ]
        
\addplot[only marks, mark=*, mark options={scale=.75,solid, fill=BrBG-5-1}]
table{%
x                      y
0.101999 0.316799
};

\addplot[only marks, mark=*, mark options={scale=.75,solid, fill=BrBG-5-2}]
table{%
x                      y
-0.058660000000000004 0.021669
-0.009076 -0.179862
-0.072572 -0.19872599999999999
0.05264 -0.283389
-0.036782999999999996 0.057561
-0.067375 -0.098523
0.030916000000000003 -0.08575
-0.23521999999999998 0.029541
0.352643 0.55787
-0.174285 -0.068388
0.16221300000000002 -0.100199
-0.23995999999999998 0.07445
-0.009835 0.051008
-0.22185700000000003 -0.015825
0.44350900000000004 0.103947
-0.129418 -0.272056
0.107028 0.17792
0.127323 0.10315
0.07435800000000001 -0.156841
-0.163516 -0.139529
0.225439 0.087228
0.051173 0.125423
-0.048554 0.219926
-0.095523 0.036821
};

\addplot[only marks, mark=*, mark options={scale=.75,solid, fill=BrBG-5-3}]
table{%
x                      y
-0.197479 0.35728499999999996
0.019793 0.057921
-0.094198 0.20842
0.108394 -0.06894700000000001
-0.050541 -0.16416
0.165621 -0.238382
0.011094 0.077771
};

\addplot[only marks, mark=*, mark options={scale=.75,solid, fill=BrBG-5-4}]
table{%
x                      y
0.037426999999999995 0.100266
-0.268131 -0.051596
-0.22967100000000001 -0.11812
-0.194444 0.10926199999999998
0.223561 0.007326999999999999
-0.087636 0.099173
0.039352 0.079796
-0.029733 0.021155
};

\addplot[only marks, mark=*, mark options={scale=.75,solid, fill=BrBG-5-5}]
table{%
x                      y
0.250118 -0.068387
-0.115927 -0.006940000000000001
0.007855 0.112053
-0.123771 -0.1829
};

\addplot[only marks, mark=diamond*, mark options={scale=1.5,solid, thick,fill=BrBG-5-2}]
table{%
x                      y
-0.049185 -0.27097
0.041958999999999996 -0.029573000000000002
};

\addplot[only marks, mark=diamond*, mark options={scale=1.5,solid, thick,fill=BrBG-5-3}]
table{%
x                      y
0.11611300000000001 -0.007918000000000001
};

\addplot[only marks, mark=diamond*, mark options={scale=1.5,solid, thick,fill=BrBG-5-4}]
table{%
x                      y
0.163574 -0.159566
};

\addplot[only marks, mark=diamond*, mark options={scale=1.5,solid, thick,fill=BrBG-5-5}]
table{%
x                      y
0.089249 -0.227196
};

\end{axis}
\end{tikzpicture}}
    \subfigure[Grouped/FE]{\begin{tikzpicture}[font=\scriptsize]
    \begin{axis}[
    width=\mdswidth,
    height=\mdsheight,
    ticks=none,
    xmin=-0.356977, xmax=0.774838,
    ymin=-0.47997100000000004, ymax=0.349587
    ]
        
\addplot[only marks, mark=*, mark options={scale=.75,solid, fill=BrBG-5-1}]
table{%
x                      y
0.138465 -0.23206999999999997
};

\addplot[only marks, mark=*, mark options={scale=.75,solid, fill=BrBG-5-2}]
table{%
x                      y
-0.022753 -0.037228
-0.19809300000000002 -0.02422
-0.15992 0.12585
-0.072238 0.24958699999999998
-0.010667 -0.024794999999999998
0.11588599999999999 0.040698000000000005
-0.047258999999999995 0.011512
-0.118094 0.069768
0.674838 -0.207314
-0.089502 0.024287
-0.11448399999999999 -0.136133
-0.063297 -0.150754
0.017269 -0.019545
-0.018647 -0.21803499999999998
0.45370299999999997 0.141074
-0.138177 0.148452
0.15779200000000002 -0.041309
0.086684 -0.066883
-0.05655399999999999 0.124483
-0.256977 -0.042604
0.11747 -0.050835000000000005
0.11419800000000001 -0.015478
0.049932 0.075086
-0.085601 -0.013130000000000001
};

\addplot[only marks, mark=*, mark options={scale=.75,solid, fill=BrBG-5-3}]
table{%
x                      y
0.049023000000000004 -0.379971
0.010646 -0.02732
0.067788 -0.131827
-0.056688 -0.064092
-0.217631 -0.176552
-0.233814 0.042554
0.046285 -0.022268
};

\addplot[only marks, mark=*, mark options={scale=.75,solid, fill=BrBG-5-4}]
table{%
x                      y
0.06275399999999999 -0.014431999999999999
-0.032732 0.065749
-0.13137000000000001 -0.029301
0.016774 0.21093
0.07144299999999999 0.219352
0.055513 -0.08462599999999999
0.048199 -0.040995
-0.03462 -0.02489
};

\addplot[only marks, mark=*, mark options={scale=.75,solid, fill=BrBG-5-5}]
table{%
x                      y
-0.129892 -0.084226
-0.10786 0.037749
0.045742000000000005 -0.054842999999999996
0.152054 0.15984
};

\addplot[only marks, mark=diamond*, mark options={scale=1.5,solid, thick,fill=BrBG-5-2}]
table{%
x                      y
-0.20496799999999998 0.221504
-0.01275 0.022347
};

\addplot[only marks, mark=diamond*, mark options={scale=1.5,solid, thick,fill=BrBG-5-3}]
table{%
x                      y
0.08862 0.05839199999999999
};

\addplot[only marks, mark=diamond*, mark options={scale=1.5,solid, thick,fill=BrBG-5-4}]
table{%
x                      y
0.070509 0.15043099999999998
};

\addplot[only marks, mark=diamond*, mark options={scale=1.5,solid, thick,fill=BrBG-5-5}]
table{%
x                      y
-0.096999 0.216031
};

\end{axis}
\end{tikzpicture}}
    \subfigure[Raw/BE]{        \begin{tikzpicture}[font=\scriptsize]
            \begin{axis}[
            width=\mdswidth,
            height=\mdsheight,
            ticks=none,
            xmin=-0.368131, xmax=0.543509,
            ymin=-0.383389, ymax=0.65787
            ]
                
\addplot[only marks, mark=*, mark options={scale=.75,solid, fill=PRGn-5-1}]
table{%
x                      y
-0.22967100000000001 -0.11812
-0.129418 -0.272056
-0.194444 0.10926199999999998
0.223561 0.007326999999999999
};

\addplot[only marks, mark=*, mark options={scale=.75,solid, fill=PRGn-5-2}]
table{%
x                      y
-0.009076 -0.179862
0.037426999999999995 0.100266
-0.072572 -0.19872599999999999
0.05264 -0.283389
-0.067375 -0.098523
-0.197479 0.35728499999999996
0.352643 0.55787
-0.174285 -0.068388
0.16221300000000002 -0.100199
-0.23995999999999998 0.07445
0.44350900000000004 0.103947
-0.123771 -0.1829
0.108394 -0.06894700000000001
0.127323 0.10315
0.07435800000000001 -0.156841
-0.163516 -0.139529
0.165621 -0.238382
0.225439 0.087228
-0.048554 0.219926
};

\addplot[only marks, mark=*, mark options={scale=.75,solid, fill=PRGn-5-3}]
table{%
x                      y
-0.268131 -0.051596
0.030916000000000003 -0.08575
-0.050541 -0.16416
};

\addplot[only marks, mark=*, mark options={scale=.75,solid, fill=PRGn-5-4}]
table{%
x                      y
-0.23521999999999998 0.029541
-0.094198 0.20842
-0.087636 0.099173
};

\addplot[only marks, mark=*, mark options={scale=.75,solid, fill=PRGn-5-5}]
table{%
x                      y
-0.058660000000000004 0.021669
-0.036782999999999996 0.057561
0.250118 -0.068387
0.101999 0.316799
-0.009835 0.051008
-0.22185700000000003 -0.015825
-0.115927 -0.006940000000000001
0.019793 0.057921
0.107028 0.17792
0.007855 0.112053
0.011094 0.077771
0.051173 0.125423
0.039352 0.079796
-0.095523 0.036821
-0.029733 0.021155
};

\addplot[only marks, mark=diamond*, mark options={scale=1.5,solid, thick,fill=PRGn-5-1}]
table{%
x                      y
0.041958999999999996 -0.029573000000000002
};

\addplot[only marks, mark=diamond*, mark options={scale=1.5,solid, thick,fill=PRGn-5-2}]
table{%
x                      y
0.089249 -0.227196
0.11611300000000001 -0.007918000000000001
-0.049185 -0.27097
0.163574 -0.159566
};

\addplot[only marks, mark=diamond*, mark options={scale=1.5,solid, thick,fill=PRGn-5-3}]
table{%
x                      y
};

\addplot[only marks, mark=diamond*, mark options={scale=1.5,solid, thick,fill=PRGn-5-4}]
table{%
x                      y
};

\addplot[only marks, mark=diamond*, mark options={scale=1.5,solid, thick,fill=PRGn-5-5}]
table{%
x                      y
};

        \end{axis}
        \end{tikzpicture}}
    \subfigure[Grouped/BE]{        \begin{tikzpicture}[font=\scriptsize]
            \begin{axis}[
            width=\mdswidth,
            height=\mdsheight,
            ticks=none,
            xmin=-0.356977, xmax=0.774838,
            ymin=-0.47997100000000004, ymax=0.349587
            ]
                
\addplot[only marks, mark=*, mark options={scale=.75,solid, fill=PRGn-5-1}]
table{%
x                      y
-0.13137000000000001 -0.029301
-0.138177 0.148452
0.016774 0.21093
0.07144299999999999 0.219352
};

\addplot[only marks, mark=*, mark options={scale=.75,solid, fill=PRGn-5-2}]
table{%
x                      y
-0.19809300000000002 -0.02422
0.06275399999999999 -0.014431999999999999
-0.15992 0.12585
-0.072238 0.24958699999999998
0.11588599999999999 0.040698000000000005
0.049023000000000004 -0.379971
0.674838 -0.207314
-0.089502 0.024287
-0.11448399999999999 -0.136133
-0.063297 -0.150754
0.45370299999999997 0.141074
0.152054 0.15984
-0.056688 -0.064092
0.086684 -0.066883
-0.05655399999999999 0.124483
-0.256977 -0.042604
-0.233814 0.042554
0.11747 -0.050835000000000005
0.049932 0.075086
};

\addplot[only marks, mark=*, mark options={scale=.75,solid, fill=PRGn-5-3}]
table{%
x                      y
-0.032732 0.065749
-0.047258999999999995 0.011512
-0.217631 -0.176552
};

\addplot[only marks, mark=*, mark options={scale=.75,solid, fill=PRGn-5-4}]
table{%
x                      y
-0.118094 0.069768
0.067788 -0.131827
0.055513 -0.08462599999999999
};

\addplot[only marks, mark=*, mark options={scale=.75,solid, fill=PRGn-5-5}]
table{%
x                      y
-0.022753 -0.037228
-0.010667 -0.024794999999999998
-0.129892 -0.084226
0.138465 -0.23206999999999997
0.017269 -0.019545
-0.018647 -0.21803499999999998
-0.10786 0.037749
0.010646 -0.02732
0.15779200000000002 -0.041309
0.045742000000000005 -0.054842999999999996
0.046285 -0.022268
0.11419800000000001 -0.015478
0.048199 -0.040995
-0.085601 -0.013130000000000001
-0.03462 -0.02489
};

\addplot[only marks, mark=diamond*, mark options={scale=1.5,solid, thick,fill=PRGn-5-1}]
table{%
x                      y
-0.01275 0.022347
};

\addplot[only marks, mark=diamond*, mark options={scale=1.5,solid, thick,fill=PRGn-5-2}]
table{%
x                      y
-0.096999 0.216031
0.08862 0.05839199999999999
-0.20496799999999998 0.221504
0.070509 0.15043099999999998
};

\addplot[only marks, mark=diamond*, mark options={scale=1.5,solid, thick,fill=PRGn-5-3}]
table{%
x                      y
};

\addplot[only marks, mark=diamond*, mark options={scale=1.5,solid, thick,fill=PRGn-5-4}]
table{%
x                      y
};

\addplot[only marks, mark=diamond*, mark options={scale=1.5,solid, thick,fill=PRGn-5-5}]
table{%
x                      y
};

        \end{axis}
        \end{tikzpicture}}\vspacePreCap
    \caption{Visualisation of CMs on ASVspoof 2019 PA task; single systems. Frontends (FEs) and backends (BEs) are compared for raw and grouped scores. Legend as in Fig.~\ref{fig:mds-cm-la}.\vspacePostCap}
    \label{fig:mds-cm-pa}
\end{figure}
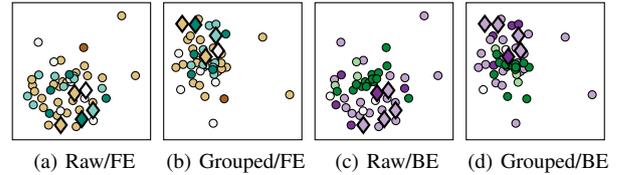

Figs.~\ref{fig:mds-cm-la} and~\ref{fig:mds-cm-pa} show visualisations for the LA and PA scenarios, respectively. Note that the location of points in each panel for each dataset (left-to-right) are identical. In each panel, circles indicate a CM, whereas diamonds signify top-5 CMs. 
Colours indicate performance and frontend/backend meta-data.

The plots give an indication of the diversity among systems. Differences between the top-5 systems are reasonably representative of the full set of systems. The diversity may imply that the top-performing single systems are complementary. This hypothesis is supported by fusion results discussed in~\cite{Nautsch2021-ASVspoof19-countermeasures} which shows that the combination of top-performing single LA and PA systems leads to improved performance. One can presume that the reason for the ineffective fusion strategies 
for PA primary system submissions might be due to lack of complementary of the systems involved in the ensemble. 

\setvariable{\mdswidth}{3.4cm}
\setvariable{\mdsheight}{3.4cm}

\subsection{Adjacency from raw vs.\ grouped scores}
Finally, we address classifier adjacency through score averaging by groups and subsequent use of the proposed method. Figs.~\ref{fig:asv-mds} and~\ref{fig:mds-cm-pa} illustrate a comparison of the proposed method for raw and grouped scores. Classifier adjacency is viewed through an alternative lens. In the ASV task, a clear separation between the two training is apparent. The VoxCeleb-trained classifiers (comparatively in-domain) become more alike. By using raw scores, trajectories by number of components (design parameters) are more alike. Feature resolution (the number of filters) corresponds to clear trajectories. In the uncontrolled CM setting, frontends as well as backends are more alike after grouping: the separation of non-GMM and GMM classifiers becomes  oblivious. 

Originally, we had two motivations for the grouped approach. First, the highly-compressed per-group average was hypothesized to give potentially a stable high-level classifier signature. Second, there is a computational advantage for very large trial sets. However, one needs additional metadata (and domain knowledge) to define the trial grouping. As we see, raw vs. grouped scores produce different findings. We prefer not to recommend either variant without a further study. A related approach, convex optimisation of error trade-offs\footnote{
    In preliminary experiments, we used the proposed method in combination with the \emph{pool adjacent violators} algorithm \cite{brummer09PAVDemonstration}. System dependently, 37720 ASV scores are reduced to 61 to 98 groups of same rank: visual differences are minimal; we spare comparisons.}, 
resonates with the Bayesian decision framework; see use of the ROC convex hull instead of the corresponding `steppy' profile~\cite{Brummer-deVilliers-BOSARIS-Binary-Scores-AGNITIO-Research-2011}.

\section{Conclusions}
We proposed a simple approach to visualise classifier adjacency on common data. An example application under controlled conditions show that the tool renders classifiers adjacent, when they are similar in metadata. In the absence of unified metadata, analyses for uncontrolled, challenge conditions prove more challenging. The proposed tool nonetheless reveals intriguing, new visual insights into classifier adjacency not provided by any existing tools.

\section{Acknowledgements}
This  work was supported  by a number of projects and funding sources:  VoicePersonae, supported by the French Agence Nationale de la Recherche~(ANR) and the Japan Science and Technology Agency (JST) with grant No.\ JPMJCR18A6; Academy of Finland (proj. 309629); Region Grand Est, France.

\bibliographystyle{IEEEtran}
\bibliography{sample}

\end{document}